\documentclass[epj,twocolumn]{webofc}
\usepackage[varg]{txfonts} 
\usepackage{comment} 
\def\eg{{\it e.g.}}
\woctitle{XLIV International Symposium on Multiparticle Dynamics}
\begin{document}
\title{Double parton correlations in Light-Front constituent quark models}
%
%

\author{Matteo
Rinaldi\inst{1}\fnsep\thanks{\email{matteo.rinaldi@pg.infn.it}} \and
        Sergio
Scopetta\inst{2}\fnsep\thanks{\email{sergio.scopetta@pg.infn.it}} \and
        Marco
Traini\inst{3}\fnsep\thanks{\email{traini@science.unitn.it}}\and
        Vicente Vento\inst{4}\fnsep\thanks{\email{vicente.vento@uv.es}}}

\institute{Dipartimento di Fisica e Geologia, Universit\`a degli Studi di
Perugia, and INFN, sezione di Perugia, via A. Pascoli 06100 Perugia, Italy
\and
           Dipartimento di Fisica e Geologia, Universit\`a degli Studi di
Perugia, and INFN, sezione di Perugia, via A. Pascoli 06100 Perugia, Italy 
\and
Dipartimento di Fisica, Universit\`a degli studi di Trento, and INFN-TIFPA,
via Sommarive 14, I - 38123 Povo (Trento), Italy      
\and
 Departament de Fisica Te\`orica, Universitat de Val\`encia
and Institut de Fisica Corpuscular, Consejo Superior de Investigaciones
Cient\'{\i}ficas, 46100 Burjassot (Val\`encia), Spain 
}

\abstract{%
  Double parton distribution functions (dPDF) represent a tool to explore
the 3D proton structure. They can be measured in
high energy
proton-proton and proton nucleus collisions and
encode information on how partons inside a proton are correlated among each
other. dPFDs are studied here
in the valence quark region, by
means of a constituent quark model, where two particle
correlations are present without any additional
prescription. This framework allows to understand the
dynamical origin of the correlations and to clarify
which, among the features of the results, are model independent.
Use will be made of
a relativistic light-front scheme, able to overcome some drawbacks of the
previous calculation. Transverse
momentum correlations, due to the exact treatment of the boosts, 
are predicted and analyzed. The role of spin
correlations is also shown. Due to the covariance of the approach, 
some symmetries of the dPDFs are seen unambigously. For the valence sector,
also the study of the QCD evolution of the
model results, which can be performed safely thanks to the
property of good support, has been also completed.
}
\maketitle
\section{Introduction}
\label{intro}
In the last few years, the study of the contribution
of  multiple parton interactions (MPI) to high energy hadron-hadron cross
sections has become an important issue. 
In these processes, more than one parton of a hadron can
interact with  partons of the other colliding hadron. Even if 
the MPI contribution is  
suppressed by a power of
$\Lambda^2_{QCD}/Q^2$ with respect to the single parton interaction, where
$Q$ is the center-mass energy, it has been 
already observed (see, \eg, Ref. \cite{2a}).
These events can be very important
as a background for the search of new  Physics, \eg, at the LHC.
In our analysis we focus actually 
on the possibilities offered by
double parton scattering (DPS), which can be
observed in several channels, \eg, $WW$ with dilepton productions
and double
Drell-Yan processes (see, Refs. \cite{3a,4a,5a,6a, report} for
recent 
reviews), to unveil the 3D nucleon structure.
At the LHC, DPS, already
observed some years ago \cite{16a}, represents also a
background for the Higgs production in several channels.

The DPS cross section
is written, following the seminal idea of Ref. \cite{1a}, in terms of the
dPDFs,
$F_{ij}(x_1,x_2,{\vec z}_\perp,\mu)$, which describe the 
joint probability 
of finding two partons of flavors $i,j=q, \bar q,g$ with 
longitudinal momentum fractions $x_1,x_2$ and separation 
$\vec z_\perp$ in the transverse plane inside the hadron:
\begin{eqnarray}  \label{eq:si_old}
  d \sigma  = \hskip -3mm & \dfrac{1}{S} &  \hskip -3mm \sum_{i,j,k,l}
\int\! d \vec z_\perp\, 
F_{ij}(x_1,x_2,\vec z_\perp,\mu) 
F_{kl}(x_3,x_4,\vec z_\perp,\mu)  \nonumber \\
  \quad &\times& 
  \hat \sigma_{ik}(x_1 x_3 \sqrt{s},\mu) \hat \sigma_{jl}(x_2 x_4
\sqrt{s},\mu)
\,.\end{eqnarray}

The cross sections $\hat \sigma$ refer to the hard parton scattering,
occurring at short-distance,
$S$ is a symmetry factor, present if identical particles appear in 
the final state and $\mu$ is the renormalization scale.
The latter is taken here, for
simplicity, to be the same for both partons.

It is worth to notice that usually,
for the evaluation and simulation of
the DPS contributions
to proton-proton scattering at the LHC kinematics,
the following approximations
are often adopted for the dPDFs: 

\begin{eqnarray} 
\label{app}
 \hskip -3mm F_{ij}(x_1,x_2,\vec z_\perp,\mu) &=& q_{i}(x_1,\mu)~
q_{j}(x_2,\mu)~T(\vec z_\perp,\mu) 
\\
\nonumber
&\times& \theta(1-x_1-x_2) (1-x_1-x_2)^n
~,
\end{eqnarray}
{\it i.e.}, dPDFs are assumed to be
completely factorized. In
particular the ${\vec z_\perp}$ and $x_1-x_2$ dependences 
are factorized and that the latter dependence can be expressed through
the standard one-body parton distribution functions (PDF), $q(x)$.
In other words,
possible double parton
correlations between the two interacting partons are neglected. 
Moreover, dPDFs are non
perturbative quantities in QCD and they cannot be easily evaluated from
the theory. In the case of PDFs, a useful procedure for the
estimate of dPDFs is their calculation at the
hadronic scale, $Q_0 \sim \Lambda_{QCD}$, by means of quark models.
In order to compare the obtained outcomes with future data taken at
high energy scales, $Q > Q_0$, it is then necessary to perform the
perturbative QCD (pQCD) evolution of the model calculations, 
using the dPDF evolution equations, which are known since a long time ago
\cite{23a,24a}.
The idea supporting our analysis is that,
thanks to this procedure, future data
of the DPS processes could be guided, in principle, by 
model calculations.
In particular, our interest is focused on the
understanding of the role of double parton correlations 
(DPCs) in the proton, in
order to verify, as a first step, 
the validity of the approximations Eq. (\ref{app}),
often used for simulations of data analyses. 
Moreover, DPCs effects,
cannot be neglected in principle (see Refs. \cite{DK,KM}
for recent contribution on this subject); 
however, DPCs are non perturbative
effects in QCD and
they can not be easily evaluated from the theory. In order to
understand the role of DPCs in the proton structure and in the
dPDFs, the use
of quark models, where DPCs are naturally included
into the scheme, can be very useful. Thanks to this
procedure, in the kinematical regions where these predictions are
reliable, the assumptions Eq. (\ref{app}) can be properly tested. 

The first model evaluations of dPDFs have been the
ones in Refs. \cite{33a,36a}. In the first scenario use has been made of a
the MIT bag model in the cavity-approximation, properly modified in order
to introduce  double parton correlations. In the second case  dPDFs have
been calculated in
a non relativistic (NR) constituent quark model (CQM) framework, since
CQMs, in the valence region, 
reproduce the gross features of PDF data and  give reasonable predictions
of generalized parton distribution functions (GPDs) and transverse momentum
dependent parton distributions (TMDs)  (see, \eg,
Refs. \cite{37a,40a,41a}).
These
expectations motivated the
analysis of Refs. \cite{36a, nostro}. The main results 
found
in Refs.  \cite{33a,36a} are that, in the valence quark region, the
approximations described in Eq. (\ref{app}) are badly violated.  
In this CQM picture, where dynamical
double correlations are present into the scheme,  the
origin of this violation can be properly understood.

One should notice that both the analyses
of Refs. \cite{33a,36a} have inconsistencies. 
First, they predict a wrong support, 
{\it i.e.},
dPDFs do not vanish in the non physical
region, $x_1+x_2 > 1$.
Moreover, as already pointed out, in order to
obtain some information on  dPDFs 
where LHC data will be taken, i.e.,
at small values of $x$ and
at high $Q^2$, the pQCD
evolution of the calculated dPDFs is necessary.
In a recent paper \cite{nostro}, we calculated dPDFs
including relativity through
a fully Poincar\'e covariant Light-Front (LF) approach. Thanks to
this treatment it is possible in general
to study strongly interacting systems with a
fixed number of on-shell constituents
(see Refs. \cite{46a,47a} for comprehensive reviews).
In this framework,
which has been extensively used  for calculations 
of hadron structure observables
(see, \eg, Refs.
\cite{48a,49a,50a}), some symmetries of  dPDFs are correctly restored
and the  bad support problem is fixed. This helps in turn to have
a precise pQCD evolution of the dPDFs.
The results of this analysis will
be summarized in the following sections.

\section{Light-Front CQM and dPDFs}
In this section, details of the calculations of dPDFs,
within the LF approach will be introduced.  In particular, the factorized
ansatz Eq. (\ref{app}), adopted for the quantities under scrutiny here,
will be studied in a relativistic scenario.
In order to have a reliable description of
dPDFs, defined in the non  perturbative region of QCD, use has been
made of the Light-Front approach, which is  well suited, as already said,
for studies of hadronic processes at high momentum transfer. 
Among all the useful
properties of this approach, let us summarize 
the ones which are appealing for our aim. 
First of all, one can
obtain a fully Poincar\'e covariant description of relativistic
strongly interacting systems with a fixed number of on-shell constituents.
Moreover, LF boosts and the ``plus'' components of momenta ($a^+ = a_0 +
a_3$) are kinematical operators and, being the LF hypersurface, the one where
the initial conditions of the system are fixed, tangent to the light-cone,
the kinematics of DIS processes is obtained naturally.

In the present analysis, the main quantity of interest is the   
Fourier- transform of the dPDF defined in Eq. (\ref{eq:si_old}): 

\begin{eqnarray} 
\label{ft}
\hskip -9mm F_{ij}^{\lambda_1,\lambda_2}(x_1,x_2,{\vec k}_\perp) 
= \int d \vec z_\perp \, e^{i \vec z_\perp \cdot \vec k_\perp} 
F_{ij}^{\lambda_1,\lambda_2}(x_1,x_2,{\vec z}_\perp)~.
\end{eqnarray}
The dPDF here introduced depends on the parton
 flavors $i$ and $j$ and on their helicities $\lambda_{i(j)}$, respectively.
Since dPDFs will be calculated here by means of a CQM, it
is necessary to find useful expressions for the dPDFs, 
by means of the LF approach. 
One needs therefore to start from the light-cone correlator which formally
defines the dPDF in quantum field theory \cite{5a}:

\begin{eqnarray}
\nonumber
\label{vera} 
\hskip -9mm  F_{ij}^{\lambda_1,\lambda_2}(x_1,x_2,{\vec z}_\perp) & = &
(-4 \pi P^{+}) \underset{\lambda}\sum \int\! d 
\vec z_\perp \, e^{i \vec z_\perp
\cdot \vec k_\perp} \int \left [ \underset{l}{\overset{3}\prod}  \dfrac{
d z_l^-}{4 \pi} \right]
\\
\nonumber
& \times &
e^{i x_1 P^+ z_1^-/2}e^{i x_2 P^+ z_2^-/2}e^{-i x_1 P^+ z_3^-/2}
\\
\nonumber
& \times &
\langle \lambda, \vec P = \vec 0  \left |
\hat{\mathcal{O}}_i^1 \left( z_1^-\dfrac{\bar n}{2},z_3^- \dfrac{\bar n}{2}+
\vec z_\perp \right) \right.
\\
& \times &
 \left. \hat{\mathcal O}_j^2 \left( z_2^-\dfrac{\bar n}{2}+
\vec z_\perp,0 \right)
\right |  \vec P = \vec 0,\lambda \rangle~,
\end{eqnarray}

where, for generic 4-vectors $z$ and $z'$, the
operator

\begin{eqnarray}
\hat{\mathcal O}_i^k(z,z') = \bar q_i(z)~ \dfrac{\bar n   \hskip-0.2cm
/}{2}  \dfrac{1 + \lambda_k \gamma_5}{2}~q_i(z')
\end{eqnarray}

has been defined in coordinate space through LF quantized fields 
of the free quarks.
In the above equation, use has been made of the light-like four vector,
$\bar n  = (1,0,0,-1)$, and of
the rest frame state of the nucleon with helicity
$\lambda$,
$\big| \vec P = \vec 0,\lambda \rangle$. 
By properly extending the procedure described in
Refs. \cite{50a,51a} for the calculation of GPDs, one can relate
dPDFs to the proton wave function, which can be calculated 
in turn through CQM.
Details on this calculation scheme can be found in Ref. \cite{nostro}. 
In particular, a crucial point of this analysis is the description of the
proton state in terms of a sum over partonic Fock states $ |\tilde k_i,
\lambda_i^f,
\tau_i \rangle$ of isospin $\tau_i$, the so called ``LF wave
function'' (LFWF) representation \cite{47a}. Moreover, in the present work,
only the first, valence contribution $|\vec 0, \lambda^f, val \rangle$  
has been taken into account
and higher Fock states have been neglected:

\begin{eqnarray}
\label{f-nstate}
\hskip -9mm |\vec 0, \lambda \rangle \simeq 
|\vec 0, \lambda^f, val \rangle &=& \underset{\lambda_i^f \tau_i}\sum
\int\left[ \underset{i=1}{\overset{3}\prod}  \dfrac{d x_i}{\sqrt{x_i}}   
\right]
\delta \left( 1- \underset{i=1}{\overset{3}\sum}x_i \right)
\\
\nonumber
&\times&
\left[
\underset{i=1}{\overset{3}\prod}  \dfrac{d \vec
k_{i\perp}}{2(2\pi)^3}  \right]2(2\pi)^3
\delta \left(  \underset{i=1}{\overset{3}\sum}\vec k_{i\perp} \right)
\\
\nonumber
&\times&\psi^{[f]}_{\lambda}(\lbrace x_i, \vec k_{i\perp} ,\lambda_i^f,
\tau_i \rbrace) \underset{i=1}{\overset{3}\prod} |\tilde k_i, \lambda_i^f,
\tau_i \rangle~,
\end{eqnarray}

where $\psi^{[f]}$ is the Light-Front proton wave function, in principle
solution of
a LF Hamiltonian directly related to the QCD lagrangian and  the short
notation $\{ \alpha_i \}$ instead of
$\alpha_1, \alpha_2, \alpha_3$ is introduced. This
representation of the proton state can be formulated in the canonical
Instant-Form as well as in the LF case through the introduction of 
canonical quark states $
|\vec k_i, \lambda_i^c,
\tau_i \rangle$:

\begin{eqnarray}
\label{I-nstate}
\hskip -9mm |\vec 0, \lambda \rangle \simeq
|\vec 0, \lambda^c, val \rangle 
&=& \underset{\lambda_i^c \tau_i}\sum
\int\left[ \underset{i=1}{\overset{3}\prod}  d \vec k_i 
\right]\delta \left(  \underset{i=1}{\overset{3}\sum}\vec k_{i} \right)
\\
\nonumber
&\times&
\psi^{[c]}_{\lambda}(\lbrace  \vec k_{i} ,\lambda_i^c,
\tau_i \rbrace) \underset{i=1}{\overset{3}\prod} |\vec k_i, \lambda_i^c,
\tau_i \rangle~,
\end{eqnarray}
where here $\psi^{[c]}$ is,  instead, the canonical proton wave function.
Since $\psi^{[f]}$ is basically  unknown, being a solution of the QCD
equations of motion, it is worth to
construct the Poincar\'e operators in order to obtain a kinematical link
between $\psi^{[f]}$ and  $\psi^{[c]}$, being the latter quantity
calculable by means of the CQM. Formally, one can always relate  free
particles
states in the LF and in the Instant-Form approaches (see Ref. \cite{46a}):
\begin{eqnarray}
\hskip -9mm | \tilde k, \lambda^f, \tau \rangle = (2\pi)^{3/2} \sqrt{m^2 +
\vec k_i^2}
\underset{\lambda^c}\sum D^{1/2}_{\lambda^f \lambda^c}(R_{cf}(\vec k)) |
\vec
k, \lambda^c, \tau \rangle~,
\label{1pstate}
\end{eqnarray}
where the Melosh rotation, which allows to rotate the
canonical helicity into the LF spin,
\begin{eqnarray}
\hskip -9mm D^{1/2}_{\mu
\lambda}(R_{cf}(\vec k_i)) =
\langle \mu  \left|~ \dfrac{m +x_iM_0 - i \vec 
\sigma_i
\cdot (\hat z_i \times \vec k_{i\perp})}{\sqrt{(m +x_i M_0)^2+\vec
k_{i\perp}^2}}
~\right| \lambda \rangle~,
\end{eqnarray}
has been introduced  through 
$x_i = \dfrac{k_i^+}{P^+}$, the longitudinal momentum fraction
carried by the parton $i$, being $P^+$  the plus component of the proton
momentum, $M_0 = \underset{i}{\sum} \sqrt{m^2+ \vec k_i^2} $  the total
free energy mass of the partonic system and $\mu$ and $\lambda$ 
generic canonical spins.
One can use then the Bakamjian-Thomas construction of the
Poincar\'e generators in order to extend the relation, Eq. (\ref{1pstate}),
to interacting systems, such as the proton one, so that a kinematical link
between the LF and Instant-Form proton state can be found \cite{nostro, 50a}:

 \begin{eqnarray}
|\vec 0, \lambda^f, val \rangle = \sqrt{M_0}(2\pi)^{3/2} |\vec 0,
\lambda^c, val \rangle~.
\label{I-C-nstate}
\end{eqnarray}
 Substituting Eqs. (\ref{1pstate}) and (\ref{I-C-nstate}) in Eq.
(\ref{f-nstate}), one can find the
relation between the LF and the canonical proton wave functions, see Refs.
\cite{nostro, 50a}:

\begin{eqnarray}
\label{I-f-wave}
\hskip -9mm 
\psi^{[f]}_\lambda (\lbrace x_i, \vec k_{i \perp},
\lambda_i^f,
\tau_i, 
\rbrace) &=& 2(2\pi)^3 \left[\dfrac{\omega_1 \omega_2 \omega_3}{M_0 x_1 x_2
x_3}  \right]^{1/2} 
\\
\nonumber
&\times&
\underset{i=1}{\overset{3} \prod} 
 \left [
\underset{\lambda_i^c}\sum D^{*1/2}_{\lambda_i^c
\lambda_i^f}(R_{cf}(\vec k_i))
\right ]
\\
\nonumber
&\times&
 \psi^{[c]}_\lambda (\lbrace  \vec k_{i},
\lambda_i^c,
\tau_i
\rbrace).
\end{eqnarray}

This is an important result for the present study,
because it allows to formally relate 
the LF proton
wave function with a
canonical proton one. 
In principle, the LF proton wave function is a
solution of the LF QCD Hamiltonian; here we adopt a phenomenological
approach where it is obtained from a CQM, reproducing only part of the
symmetries of the theory.
Using a lengthy but straightforward procedure,
the final expression of the dPDF is obtained. It reads \cite{nostro}:
\begin{eqnarray}
\label{main}
\nonumber
\hskip -9mm  F_{q_1 q_2}^{\lambda_1,\lambda_2}(x_1, x_2, \vec k_\perp) 
& = & 
3(\sqrt{3})^3 \int
\left[
\underset{i=1}{\overset{3}\prod} d \vec k_i 
\underset{\lambda_i^f \tau_i} {\sum}
\right]
\delta \left(
\underset{i=1}{\overset{3}\sum} \vec k_i 
\right) 
\\
\nonumber
&\times&
\Psi^* \left(\vec k_1 +
\dfrac{\vec k_\perp}{2}, \vec k_2 -
\dfrac{\vec k_\perp}{2},\vec k_3
; \{\lambda_i^f, \tau_i \}
\right) \\
\nonumber
& \times & \widehat{P}_{q_1}(1)\widehat{P}_{q_2}
(2)\widehat { P } _ { \lambda_1 } (1)\widehat { P } _ { \lambda_2 } (2)
\, 
\\
\nonumber
&\times&
\Psi \left(\vec k_1 -
\dfrac{\vec k_\perp}{2}, \vec k_2 +
\dfrac{\vec k_\perp}{2}, \vec k_3 
; \{\lambda_i^f, \tau_i \}
\right)
\\
& \times & \delta \left(x_1 
-\dfrac{k_1^+}{P^+}
\right) \delta \left(x_2 -\dfrac{k_2^+}{P^+}
\right)~.
\end{eqnarray}

\begin{figure}[t]
\begin{minipage}[t] {70 mm}
\vspace{7.0cm}
\includegraphics{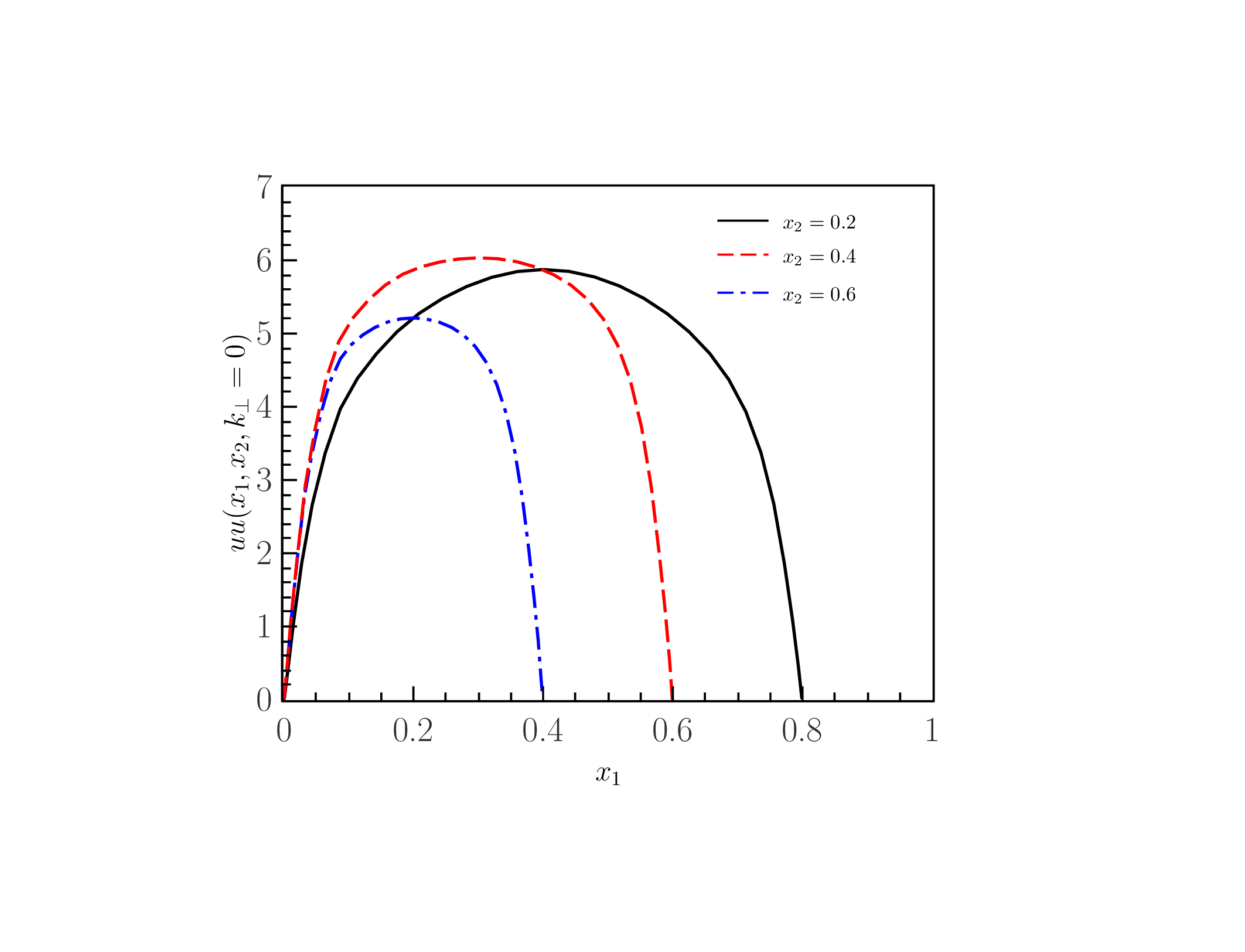}
\vskip -.8cm
\caption{ \footnotesize The distribution $uu(x_1,x_2,k_\perp)$, Eq.
(\ref{unp}), for three
values of $x_2$ and $k_\perp=0$.}
\label{1}
\end{minipage}
\hspace{\fill}
\begin{minipage}[t] {70 mm}
\vspace{7.0cm}
\includegraphics{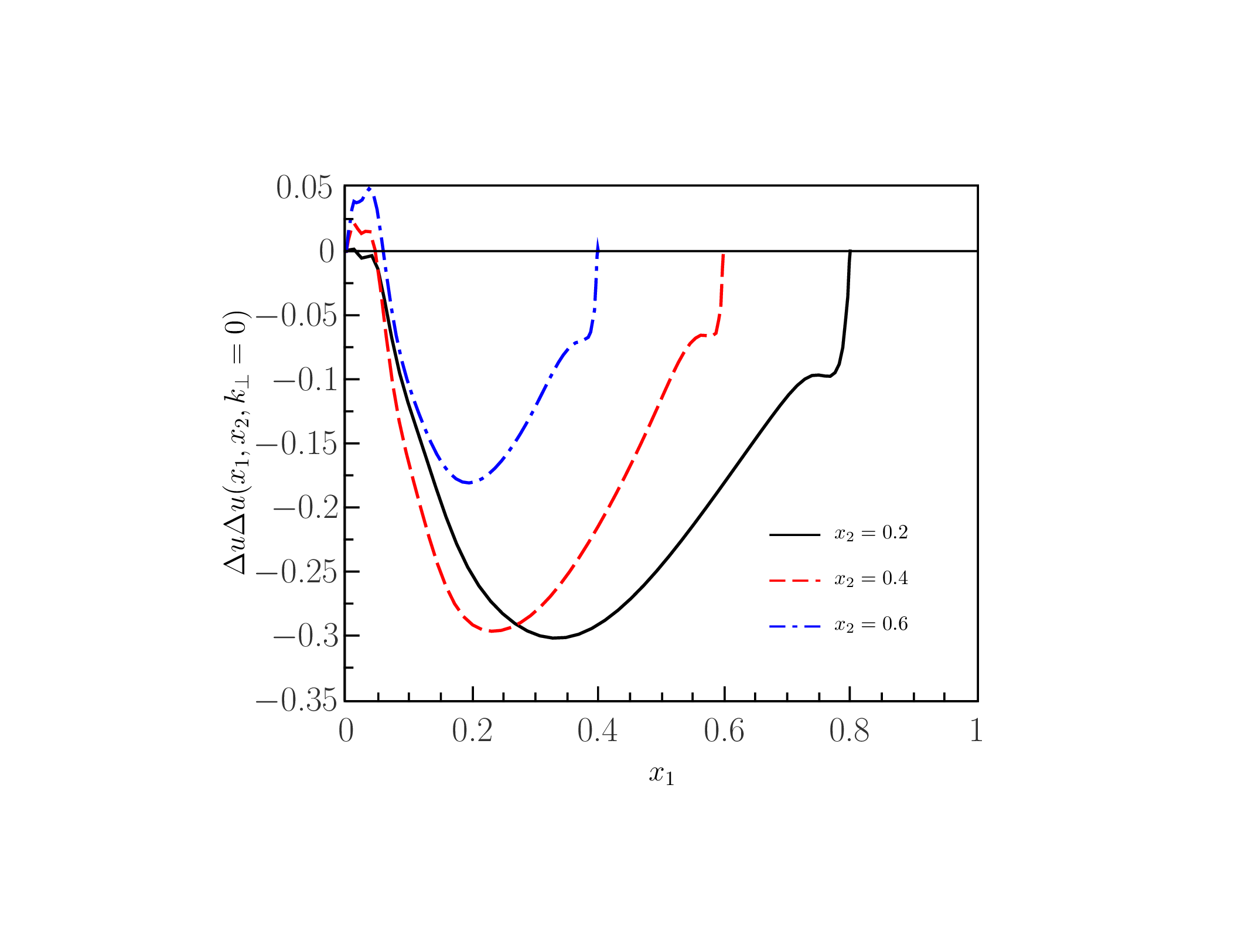}
\vskip -.2cm
\caption{\footnotesize The distribution $\Delta u \Delta u
(x_1,x_2,k_\perp)$, Eq.
(\ref{pol}),
for three
values of $x_2$ and $k_\perp=0$.}
\label{2}
\end{minipage}
\end{figure}

\begin{figure}[t]
\begin{minipage}[t] {70 mm}
\vspace{7.0cm}
\includegraphics{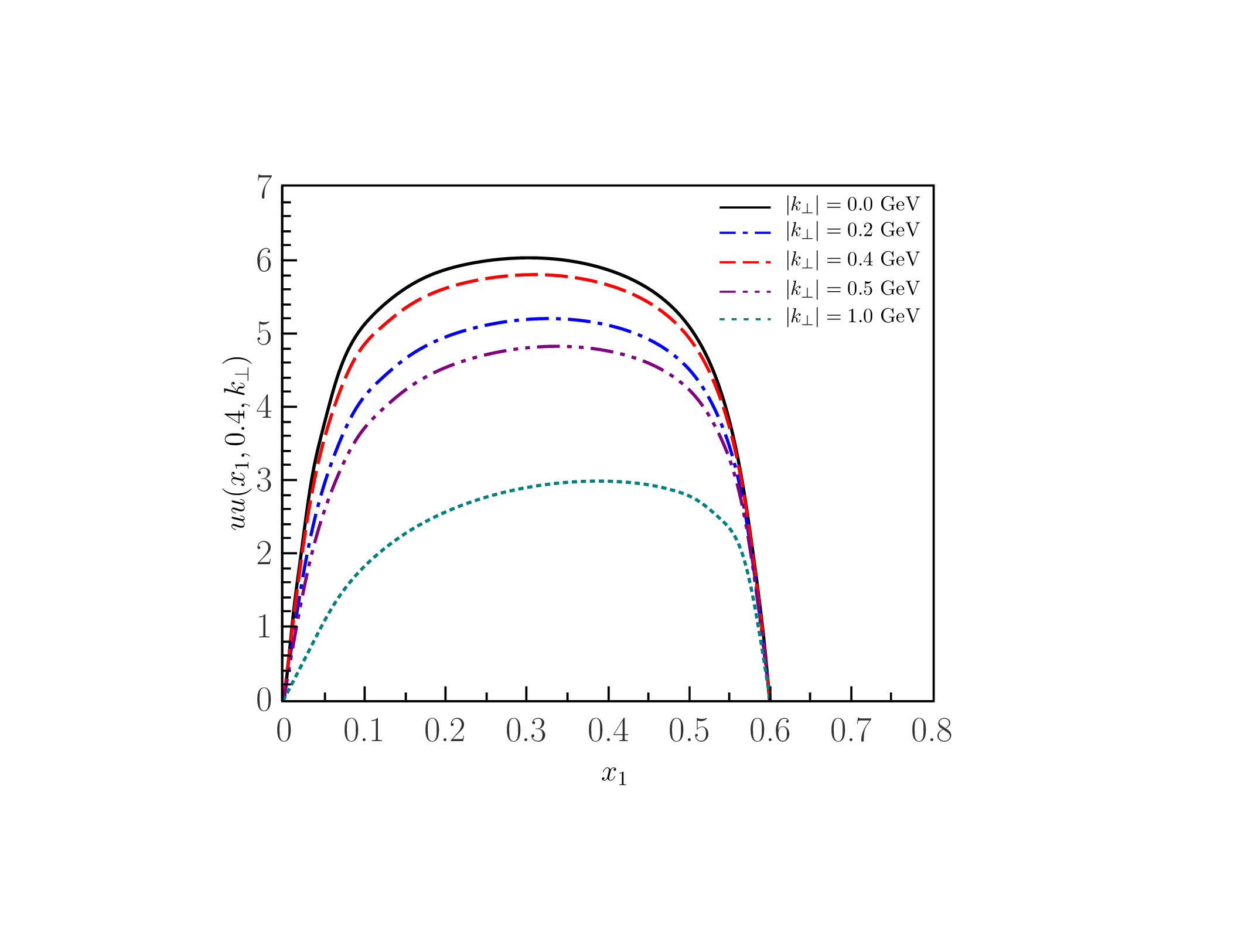}
\vskip -0.8cm
\caption{ \footnotesize The distribution $uu(x_1,x_2,k_\perp)$, Eq.
(\ref{unp}), for 
 $x_2=0.4$ and five values of $k_\perp$.}
\label{3}
\end{minipage}
\hspace{\fill}
\begin{minipage}[t] {70 mm}
\vspace{7.0cm}
\includegraphics{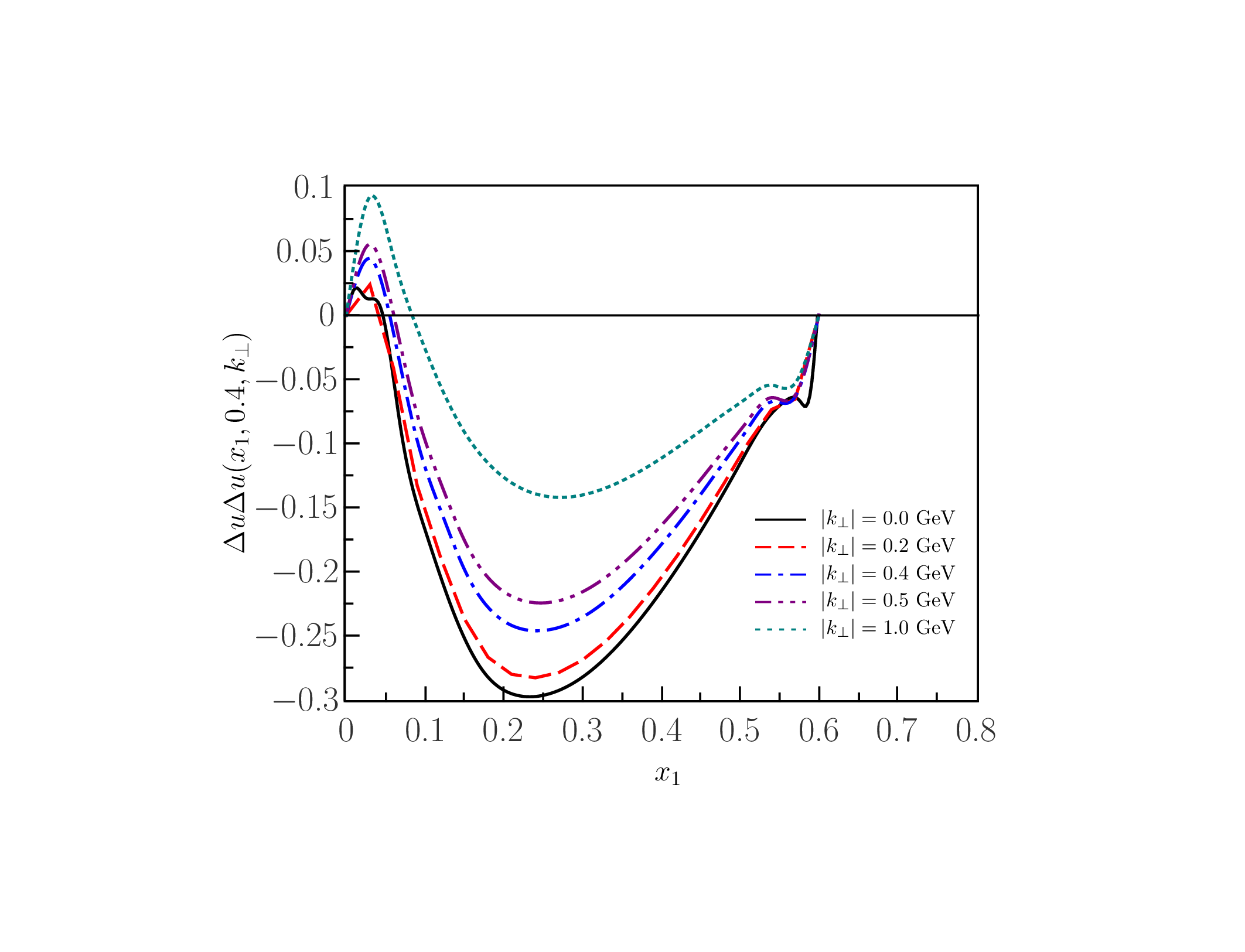}
\vskip -0.2cm
\caption{\footnotesize The distribution $\Delta u \Delta
u(x_1,x_2,k_\perp)$, Eq.
(\ref{pol}),
for 
 $x_2=0.4$ and five values of $k_\perp$.}
\label{4}
\end{minipage}
\end{figure}

\begin{figure}[t]
\begin{minipage}[t] {70 mm}
\vspace{7.0cm}
\includegraphics{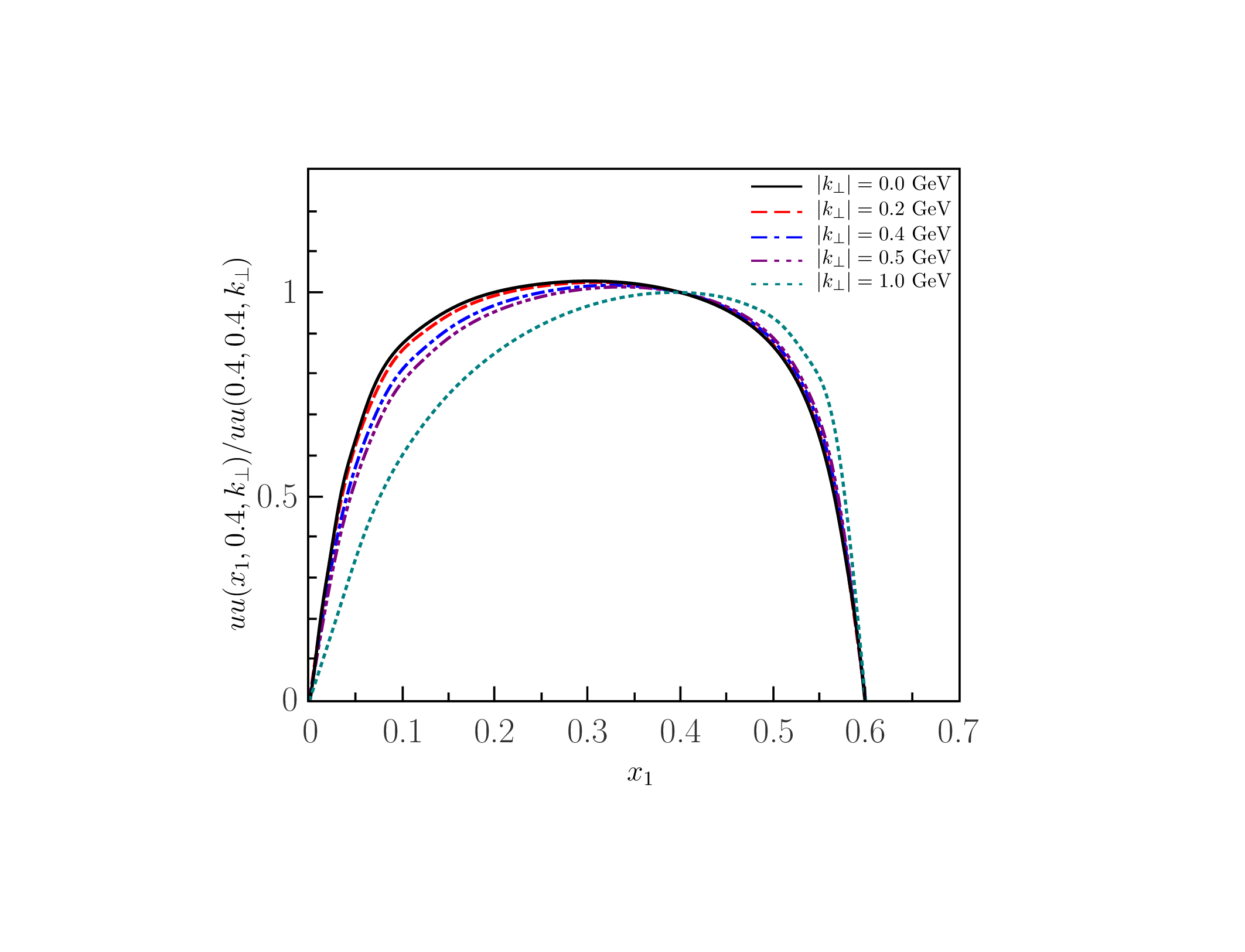}
\vskip -0.8cm
\caption{ \footnotesize The ratio $r_1$, Eq. (\ref{ratio12u}), for
five values of $k_\perp$.}
\label{5}
\end{minipage}
\hspace{\fill}
\begin{minipage}[t] {70 mm}
\vspace{7.0cm}
\includegraphics{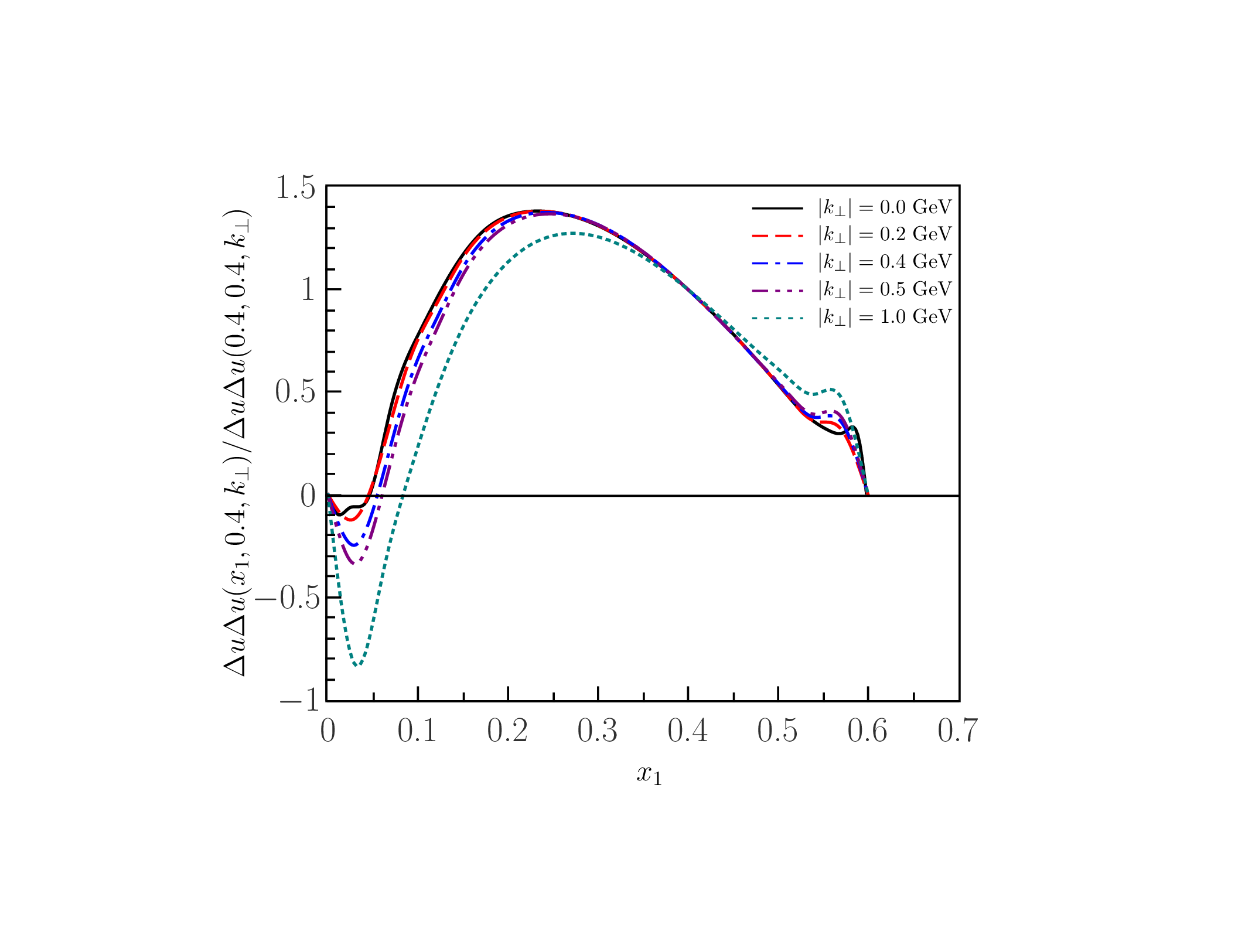}
\vskip -0.2cm
\caption{\footnotesize The ratio $r_2$, Eq. (\ref{ratio12p}), 
for five values of $k_\perp$.}
\label{6}
\end{minipage}
\end{figure}

\begin{figure}[t]
\begin{minipage}[t] {70 mm}
\vspace{7.0cm}
\includegraphics{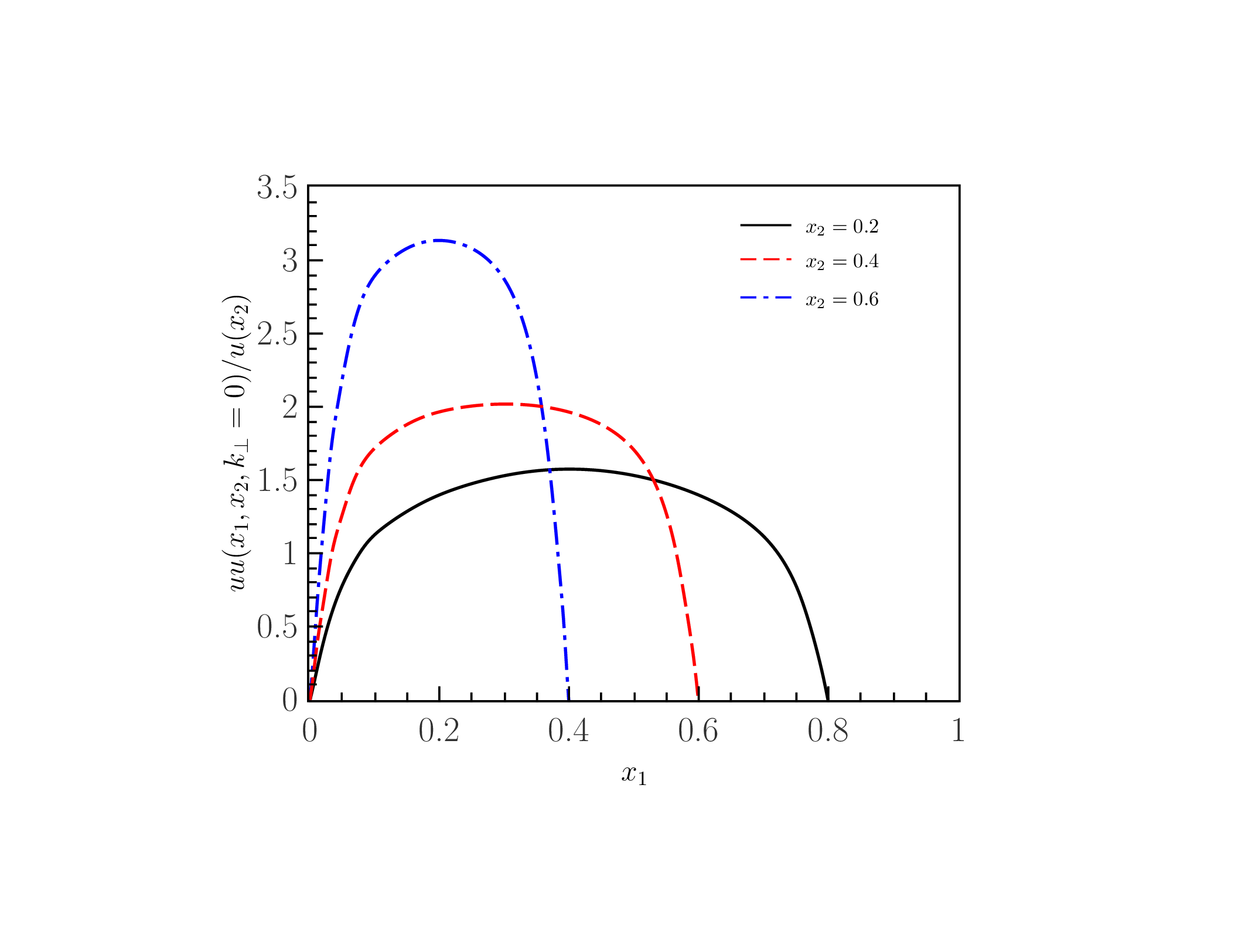}
\vskip -0.8cm
\caption{ \footnotesize The ratio $r_3$, Eq. (\ref{ratio34nu}), for three
values of $x_2$ and $k_\perp=0$.}
\label{7}
\end{minipage}
\hspace{\fill}
\begin{minipage}[t] {70 mm}
\vspace{7.0cm}
\includegraphics{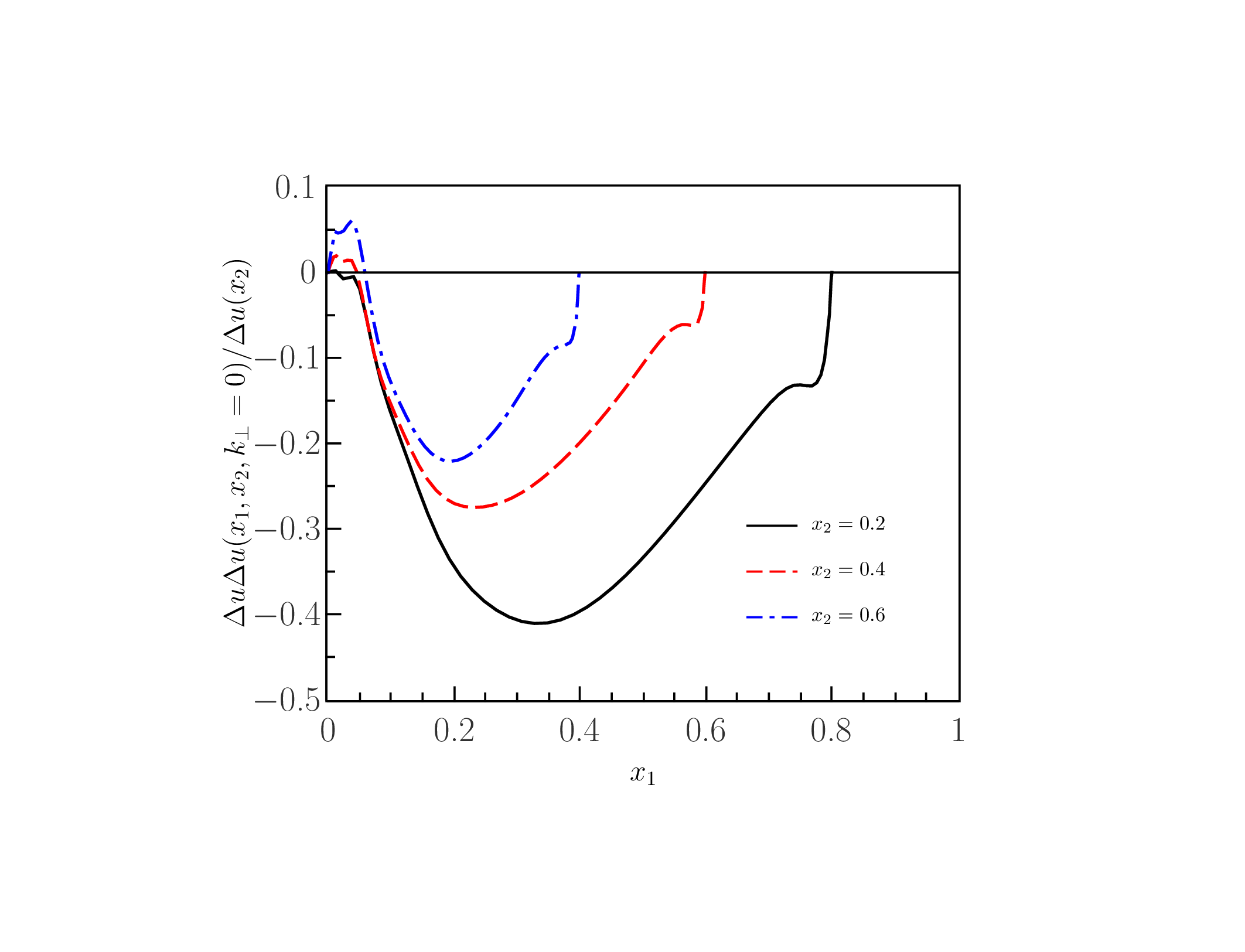}
\vskip -0.2cm
\caption{\footnotesize The ratio $r_4$, Eq. (\ref{ratio34np}), 
for three
values of $x_2$ and $k_\perp=0$.}
\label{8}
\end{minipage}
\end{figure}

The canonical proton wave function $\psi^{[c]}$ is embedded in
the function
$\Psi$ here above, which can be written as follows:

\begin{eqnarray}
\label{psiint}
\hskip -3mm \Psi(\vec k_1,\vec k_2,\vec k_2; \lbrace \lambda_i^f, \tau_i
\rbrace) &=&
\underset{i=1}{\overset{3} \prod} \left [
\underset{\lambda_i^c}\sum D^{*1/2}_{\lambda_i^c
\lambda_i^f}(R_{cf}(\vec k_i))
\right ]
\\
\nonumber
&\times&
 \psi^{[c]}(\{\vec k_i, \lambda_i^c, \tau_i   \} )~,
\end{eqnarray}
where $\lambda_i^c$ and $\tau_i$ are the canonical parton helicity and
the isospin, respectively. Isospin and spin
projection operators are introduced in order to define  dynamical
correlations for the unpolarized and longitudinal polarized $i$ quark of a
given flavor, as follows:

\begin{eqnarray}
\label{proj}
\hat P_{u(d)}(i) = \dfrac{ 1 \pm \tau_3(i)}{2}~, ~
\hat P_{\lambda_k}(i)  =  \dfrac{ 1 + \lambda_k \sigma_3(i)}{2}~.
\end{eqnarray}
In this work the use of the LF approach is crucial  since
the plus component of the momenta is kinematical so that, in the intrinsic
frame, where $\vec k_1 + \vec k_2+ \vec k_3=0$, one finds the following
condition:

\begin{eqnarray}
P^+= \overset{3}{ \underset{i}\sum} k_i^+  = M_0~. 
\end{eqnarray}
Thanks to the latter relation, the delta function, defining the longitudinal
momentum fractions carried by the partons in Eq. (\ref{main}), can be
easily worked out without any additional approximation, at variance with
what happens in the instant form 
calculations of PDFs and  dPDFs (see, \eg, Ref. \cite{36a}). 
As a direct consequence, the bad support
problem does not show up in our relativistic calculation.
Let us remark again that the main ingredient in Eq. (\ref{main})
is the canonical proton wave function so that, in order to estimate dPDFs,
verifying if the approximations, Eq. (\ref{app}), hold, use has been made of a
relativistic CQM, the so called hyper-central CQM described in Ref. \cite{49a}.
For the present analysis it is sufficient to know that the
proton wave function can be factorized in terms of the momentum space
function and the spin-isospin one. As a first approximation,
the latter part will be described
within SU(6) symmetry. Moreover, the momentum space function is
obtained by solving a relativistic effective Mass equation, 
described through a potential model in agreement with the
Bakamjian-Thomas construction \cite{nostro,49a}. 
The choice
of this model is motivated by its simplicity and capability to basically
reproduce  the spectrum of light-hadrons. 
Indeed it has been already used
for the estimate  of PDFs and GPDs, see Refs.  \cite{49a,
50a,51a,54a,55a}. Moreover, in the dPDFs case, there are not yet data available
and model calculations of these quantities could give
essential information on their relevant features.

\section{Results of the calculations at the hadronic scale}

In this section the main results, obtained at the hadronic energy scale,
$\mu_0^2 \sim 0.1$ GeV$^2$, will
be discussed. In particular, the emphasis of the present analysis is focused on
testing the validity of the approximations Eq. (\ref{app}). 
To this aim, in this work, two
specific combinations of the components of 
the spin-dependent dPDFs, Eq. (\ref{main}), for
two quarks of flavor $u$, have been
evaluated: 

\begin{eqnarray}
 \label{unp}
uu(x_1,x_2, k_\perp) &=& \underset{i,j = \uparrow,\downarrow}\sum
u_iu_j(x_1,x_2, k_\perp),
\\
\label{pol}
\Delta u \Delta u(x_1,x_2, k_\perp) &=& \underset{i=
\uparrow,\downarrow}\sum
u_iu_i(x_1,x_2, k_\perp)
\\
\nonumber
&-&\underset{i\neq j =
\uparrow,\downarrow}\sum
u_iu_j(x_1,x_2, k_\perp)~,
\end{eqnarray}
being these distributions observable, in principle, in processes involving
unpolarized protons. 
In Figs. \ref{1} and 2 the distributions introduced in Eqs. (\ref{unp},
\ref{pol})
are shown for three values of $x_2$ and for $k_\perp=0$. One
can easily realize that the support problem has been solved, {\it i.e.},
the dPDFs are different from zero only in physical kinematical regions,
when $x_1+x_2 \leq 1$. Moreover, thanks to the overcome of the
inconsistency present in the canonical calculations of dPDFs in
Ref. \cite{33a}, the symmetry in exchanging $x_1$ and $x_2$, due to
the indistinguishability of the two interacting particles when $k_\perp=0$,
is restored.

In Figs. 3 and 4 the same distributions are shown for five values
of $k_\perp$ and  for $x_2=0.4$. In particular, the decreasing behavior of
the dPDFs with respect to the increasing of $k_\perp$ 
is qualitatively similar to the results discussed in Refs. \cite{33a,36a}. 
In order  to study the validity of the approximations Eq.
(\ref{app}), in particular the factorization of dPDFs in the $k_\perp$
and $x_1, x_2$ dependence, the following ratios have been evaluated:

\begin{eqnarray}
\label{ratio12u}
\hskip -9mm  r_1 &=& \dfrac{uu(x_1,0.4,k_\perp )}{uu(0.4,0.4,k_\perp )},
\\
\nonumber
\\
\label{ratio12p}
r_2
&=&
\dfrac{\Delta u \Delta u(x_1,0.4,k_\perp )}{\Delta u \Delta
u(0.4,0.4,k_\perp )}~.
\end{eqnarray}

In Figs. 5 and 6 the calculated ratios $r_1$ and $r_2$ have been shown
for five
values $k_\perp$. As one can see, these quantities vary weakly on 
$k_\perp$  so that a factorized ansatz of dPDFs on $k_\perp$ and $x$
dependence is not valid in this approach, as in the cases discussed
in Refs. \cite{33a,36a}. Let us remark in particular that the amount
of this violation is strongly related to the contribution of the Melosh
rotations, present also in the unpolarized  case when $k_\perp \neq 0$.
These kind of relativistic effects are model independent.

In order to
study the factorization on the $x_1$ and the $x_2$ dependences, the
following ratios have been introduced:

\begin{eqnarray}
\label{ratio34nu}
 r_3 &=& \dfrac{ uu(x_1,x_2, k_\perp=0)}{u(x_2)}~,
 \\
 \nonumber
\\
\label{ratio34np}
 r_4 &=& \dfrac{\Delta u \Delta u(x_1,x_2, k_\perp=0)}{
\Delta u(x_2)}~,
\end{eqnarray}

where the standard PDFs, calculated within the same
hyper-central CQM,  $u(x)$ and  $\Delta u(x)$, in the unpolarized and
polarized cases, respectively, have been
introduced to analyze if  dPDFs can be factorized, in terms of the
product of two single PDFs,  in this relativistic approach. 
In Figs. 7 and 8
the ratios $r_3$ and $r_4$, calculated for
three values of $x_2$ and for $k_\perp=0$, are shown.  As one
can see, these quantities
depend on $x_2$, at variance with what would happen if the approximation,
Eq.
(\ref{app}), were valid. One can deduce therefore that the dPDFs here
calculated are strongly different form the product 
two single PDFs. These results are
in qualitatively agreement with
the ones of Refs. \cite{33a,36a}.
To have 
more details on the violation on the factorization
of longitudinal momentum dependences, 
the following  ratios have been introduced and
calculated:

\begin{eqnarray}
\label{ratio34u}
 r_5 &=& \dfrac{2 uu(x_1,x_2, k_\perp=0)}{u(x_1)u(x_2)}~,
 \\
 \nonumber
\\
\label{ratio34p}
 r_6 &=& \dfrac{C \Delta u \Delta u(x_1,x_2, k_\perp=0)}{\Delta u(x_1)
\Delta u(x_2)}~,
\end{eqnarray}
where:
\begin{eqnarray}
 C = \dfrac{ [\int dx~ \Delta u(x)]^2  }{\int d x_1 d x_2~ \Delta u
\Delta u(x_1,x_2,k_\perp =0)}~.
\end{eqnarray}
The single PDFs appear in the denominators of
the above equations. Besides, the factors $2$ and $C$ are properly
inserted in,
Eqs. (\ref{ratio34u}) and (\ref{ratio34p}), in order to normalize these
ratios in the kinematical
regions where the correlations, in the longitudinal momenta, can be
neglected, in
agreement with the approximation of Eq. (\ref{app}).

\begin{figure}[t]
\vspace{15.0cm}
\includegraphics{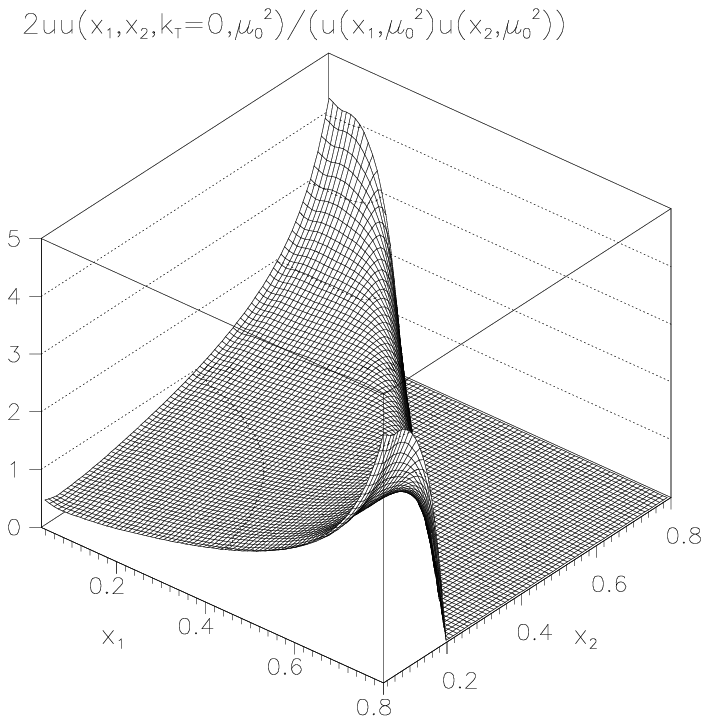}
\vskip -8cm
\caption{ \footnotesize  The ratio $r_5$, Eq. (\ref{ratio34u}), at the
hadronic scale.} 
\label{9}
\end{figure}

\begin{figure}[t]
\vspace{15.0cm}
\includegraphics{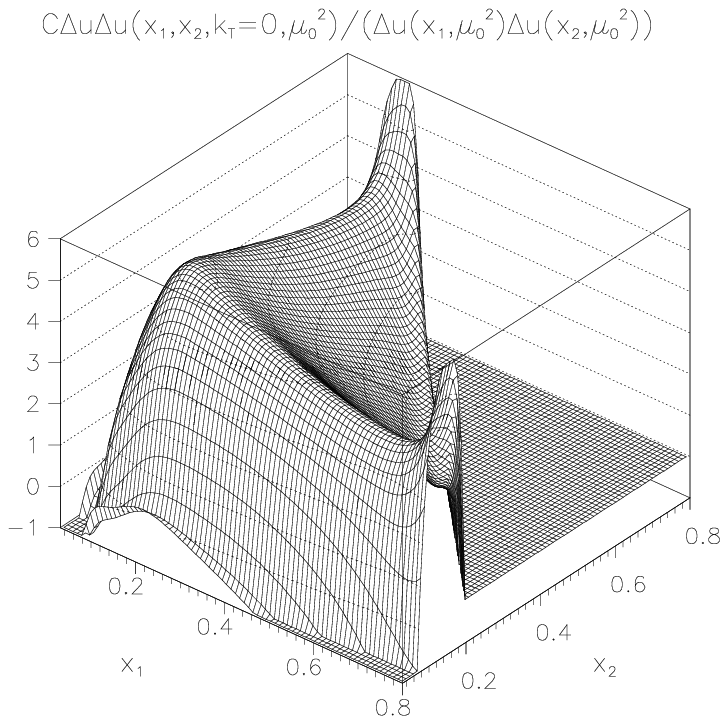}
\vskip -9cm
\caption{ \footnotesize 
 the ratio $r_6$, Eq. (\ref{ratio34p}), at the hadronic scale $\mu_0$.} 
\label{10}
\end{figure}

\begin{figure}[t]
\vspace{15.0cm}
\includegraphics{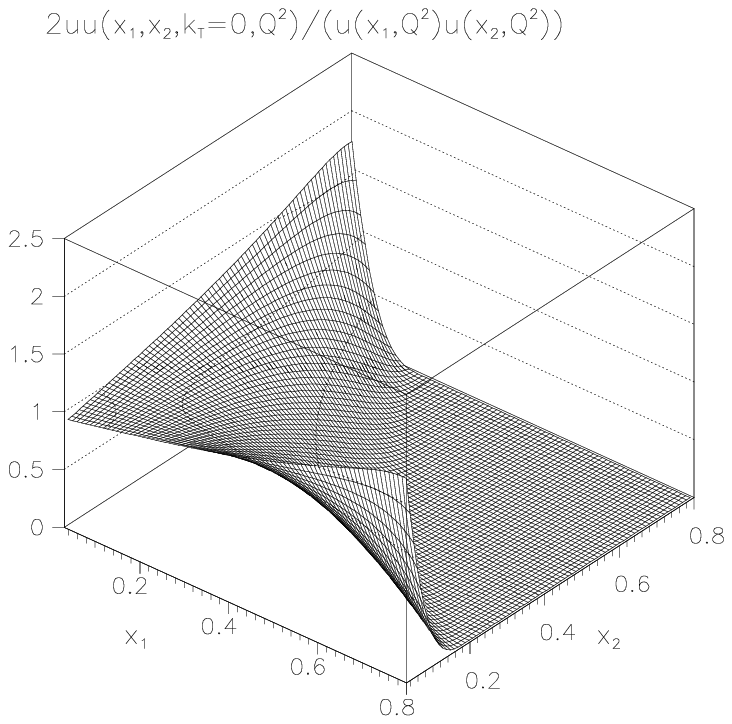}
\vskip -7.5cm
\caption{ \footnotesize 
The ratio $r_5$,  Eq. (\ref{ratio34u}),  at a scale $Q^2= 10$ GeV$^2$.} 
\label{11}
\end{figure}

\begin{figure}[t]
\vspace{15.0cm}
\includegraphics{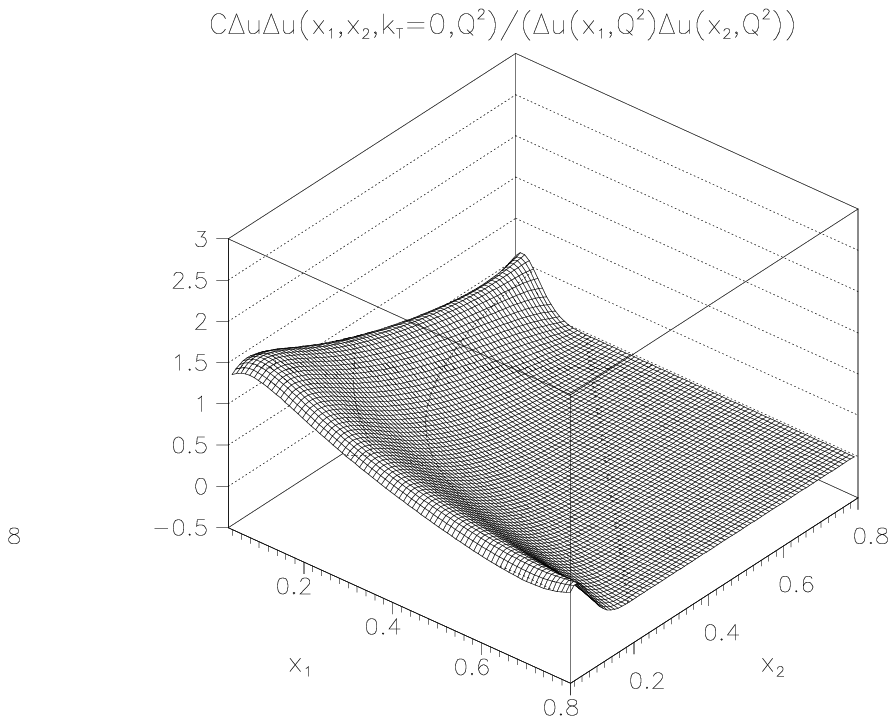}
\vskip -8.5cm
\caption{ \footnotesize 
 the ratio $r_6$, Eq. (\ref{ratio34p}), at a scale $Q^2= 10$ GeV$^2$.} 
\label{12}
\end{figure}

As one can see in Figs. 9 and 10,  a
factorized form of dPDFs in terms of single PDFs is not supported by 
the present
approach in almost all the kinematical regions where the predictions of CQM
calculations are reliable.

\section{pQCD evolution of the calculated dPDFs}

A fundamental point discussed in in Ref. \cite{nostro} is the 
analysis of the pQCD evolution of our dynamical model of dPDFs.
This
procedure is essential to relate 
CQM predictions with future data  taken  at
the LHC or with simulations of DPS process contributions  to 
events in particular channels in  $pp$ and $pA$ collisions. 
For the moment
being,  the pQCD evolution of
dPDFs is performed only in the longitudinal momentum dependence, which means
$k_\perp=0$, and using the same energy scale for both the acting
partons. In these case,  the evolution equations are obtained as a proper
generalization of the usual  Dokshitzer–Gribov–Lipatov–Altarelli–Parisi
(DGLAP)  ones (see Refs. \cite{23a,24a} for
details), defined for the evolution of PDFs. 
In the present analysis only the contribution of the valence, 
non-singlet sector, has been taken into account
in the evolution at Leading-Order, so that one
needs to solve only the homogeneous part of of the evolution equations 
by using the Mellin transformations of 
dPDFs (see Ref. \cite{nostro} for details). 
We notice that,
if we use the simple ansatz discussed in Ref. \cite{30a} for the evaluation
of dPDFs, our code for the pQCD evolution
reproduce the results found and discussed in that paper. 
After having obtained the dPDFs 
at a generic high energy scale, \eg, $Q^2 = 10$ GeV$^2$, 
using pQCD, the ratios $r_5$
and $r_6$ have been shown again in Figs. 11 and 12. The most important
results of this analysis are that, for small values of $x$, \eg, close
to the LHC kinematics,  $r_5 \sim 1$. This means that, in the
unpolarized case, dynamical correlations are suppressed after the
evolution. Nevertheless, by looking at $r_6$, in Fig 6, it is found that
double spin correlations still contribute, even at low $x$.

\section{Conclusions}
In this work, dPDFs appearing in the  DPS cross section have been
evaluated within a LF CQM. Fully
Poincar\'e 
covariance of the description, which
allows to restore the expected symmetries, and the vanishing of  dPDFs in
the forbidden kinematical region, $x_1+x_2>1$, is achieved. 
In the analysis of  dPDFs at the hadronic scale, 
the approximations of these quantities with a
complete factorized ansatz, in the $x_1-x_2$ and $(x_1,x_2)-k_\perp$
dependences, are found to be violated, in agreement with  previous results  
\cite{33a,36a}. Moreover, a pQCD analysis of the valence dPDFs, necessary to
evaluate these quantities at higher energy scales with respect to the
hadronic one where the CQM predictions are valid, has been
performed. For the unpolarized quarks case dynamical
correlations are suppressed in the small $x$ region, while double
spin correlations are found to be still important. Further analysis, 
including into the scheme non perturbative sea quarks and gluons 
together with the evolution of dPDFs including the
singlet sector contribution, fundamental in order to describe the dPDF at
low $x$, are under way, as well as the study of the extraction of
proton dPDFs from $pA$ collisions, along the line of Ref. \cite{25}.

\section{Acknowledgments}

This work was supported in part by the Research Infrastructure
Integrating Activity Study of Strongly Interacting Matter (acronym
HadronPhysic3, Grant Agreement n. 283286 and n. 283288) 
under the Seventh Framework
Programme of the European Community,
by the Mineco
under contract FPA2010-21750-C02-01,
by GVA-Prometeo/2009/129, and by CPAN(CSD-00042).
M.R. thanks the organizers of the conference for the invitation.

\end{document}